\documentclass{JHEP3}
\usepackage{amsmath,amssymb}

\def\del{\partial}
\def\ee{{\mathrm{e}}}
\def\alangle{\Big\langle\!\!\Big\langle}
\def\arangle{\Big\rangle\!\!\Big\rangle}
\def\m{\tilde{m}}
\renewcommand{\epsilon}{\varepsilon}
\newcommand{\Sp}{\text{Sp}}
\newcommand{\SU}{\text{SU}}
\newcommand{\U}{\text{U}}
\newcommand{\Z}{\mathbb{Z}}
\newcommand{\1}{\mathbf{1}}
\newcommand{\D}{\mathcal{D}}
\DeclareMathOperator{\re}{Re}
\DeclareMathOperator{\tr}{Tr}
\DeclareMathOperator{\pf}{Pf}
\DeclareMathOperator{\sgn}{sgn}

\title{Chiral Lagrangian and spectral sum rules for dense two-color QCD}
\author{Takuya Kanazawa$^a$, Tilo Wettig$^b$, and Naoki Yamamoto$^a$\\
  $^a\,$Department of Physics, The University of Tokyo, Tokyo
  113-0033, Japan\\
  \mbox{$^b\,$Department of Physics, University of Regensburg, 93040
    Regensburg, Germany}\\  
  Email: \email{tkanazawa@nt.phys.s.u-tokyo.ac.jp},
  \email{tilo.wettig@physik.uni-regensburg.de},
  \email{yamamoto@nt.phys.s.u-tokyo.ac.jp}}

\abstract{We analytically study two-color QCD with an even number of
  flavors at high baryon density.  This theory is free from the
  fermion sign problem.  Chiral symmetry is broken spontaneously by
  the diquark condensate.  Based on the symmetry breaking pattern we
  construct the low-energy effective Lagrangian for the
  Nambu-Goldstone bosons.  We identify a new epsilon-regime at high
  baryon density in which the quark mass dependence of the partition
  function can be determined exactly.  We also derive
  Leutwyler-Smilga-type spectral sum rules for the complex eigenvalues
  of the Dirac operator in terms of the fermion gap.  Our results can
  in principle be tested in lattice QCD simulations.}


\preprint{TKYNT-09-10; August 9, 2009}

\begin{document}

\section{Introduction}

Quantum chromodynamics (QCD) exhibits various phases under extreme
conditions: the quark-gluon plasma phase at high temperature $T$ and
the color superconducting phase \cite{Rajagopal2000,Alford2008} at low
$T$ and intermediate or large quark chemical potential $\mu$.  The
quark-gluon plasma phase is being studied experimentally in
ultrarelativistic heavy-ion collisions and theoretically in lattice
QCD simulations.  On the other hand, the physics of color
superconductivity at intermediate $\mu$, which is important for the
interior of neutron stars, is not yet fully understood, although many
possible phases have been proposed, e.g., the meson condensed phase,
the crystalline Fulde-Ferrell-Larkin-Ovchinikov phase, the gluon
condensed phase, etc.~\cite{Alford2008}.  The case of
asymptotically large $\mu$ is better understood because in this region
rigorous weak-coupling calculations can be performed owing to
asymptotic freedom, and the ground state has been shown to
be the most symmetric, color-flavor-locked (CFL) phase
\cite{Alford:1998mk}.

Given the enormous theoretical and phenomenological interest in QCD at
nonzero density, it is unfortunate that first-principle lattice
simulations are extremely difficult because of the fermion sign
problem: The Dirac determinant becomes complex for nonzero $\mu$, and
conventional Monte Carlo methods fail.  However, there is a category
of theories, including two-color QCD with fundamental fermions and QCD
with an arbitrary number of colors and adjoint fermions, that are free
from the sign problem because of an additional anti-unitary symmetry.
Although these theories are quite different from QCD in certain
respects (e.g., the pattern of chiral symmetry breaking is different
from that of real QCD), they also exhibit dynamical phenomena such as
diquark condensation that are expected to occur in real QCD, and
thus we hope that they can provide us with some clues about the phase
structure of real QCD. Recent years have seen substantial progress in
the understanding of these QCD-like theories at nonzero temperature
and density, both analytically and numerically
\cite{Peskin:1980gc,Hands1999,Kogut1999,Kogut2000,Hands2000,Splittorff2001,
  Vanderheyden2001,Kogut2001,
  Lenaghan2002,Splittorff2002,Splittorff2002a,Kogut2002,
  Wirstam2003,Muroya:2002ry,Schafer2003,Kogut:2003ju,Nishida2004,Klein2005,Ratti2004,
  Alles:2006ea,Akemann:2006xn,Hands2006,Alles2007,Fukushima2007}.

In particular, two-color QCD has been studied intensively. Two-color
QCD at small $\mu$ has been investigated by chiral perturbation theory
in the mean-field approximation \cite{Kogut1999,Kogut2000}, later
generalized in various directions
\cite{Splittorff2001,Splittorff2002,Splittorff2002a}, and many
intriguing phenomena have been revealed, such as the occurrence of a
second-order phase transition at $\mu = m_\pi/2$, where $m_\pi$
is the pion mass. Also, vector mesons have been included in the chiral
Lagrangian, which led to the prediction of vector meson condensation
\cite{Lenaghan2002,Sannino2002,Sannino2003}.  However,
these approaches are based on an expansion in small $\mu$ and cannot
be used to probe the system at large $\mu$.

On the other hand, at \emph{asymptotically large} $\mu$ the formation
of a BCS superfluid of diquark pairs is expected based on the
existence of an attractive channel between quarks near the Fermi
surface.  The gap for quasiparticles has been calculated
perturbatively in \cite{Son1999a, Schafer2003}.  In this case, chiral
symmetry is spontaneously broken not by the chiral condensate but by
the diquark condensate, and the pattern of chiral symmetry breaking is
quite different from that at $\mu\sim 0$.

In this paper we study two-color QCD at large $\mu$ analytically. We
construct the chiral Lagrangian for the Nambu-Goldstone (NG) bosons
associated with the spontaneous breaking of chiral symmetry for the
case of an even number of flavors and determine the form of the mass
term by symmetries.\footnote{See
  \cite{Casalbuoni1999e,Son2000,Son2000a,Schafer2002} for a similar
  construction in the CFL phase of three-color QCD.}  Furthermore we
consider the system on a finite torus and identify the corresponding
``$\epsilon$-regime'' (or ``microscopic domain'') of the theory in which
the finite-volume physics is completely dominated by the zero-momentum
modes of the NG bosons.  In this regime we can calculate the
finite-volume partition function analytically as a function of the
light quark masses.  From this result we derive a set of novel
spectral sum rules for inverse powers of the Dirac eigenvalues. These
sum rules are a generalization of the well-known Leutwyler-Smilga sum
rules at $\mu=0$ \cite{Leutwyler1992}, but our analysis is more
rigorous in the sense that the occurrence of spontaneous symmetry
breaking is not assumed, but derived, from the microscopic dynamics of
QCD based on asymptotic freedom.  We also propose that a microscopic
spectral density can be defined similarly to the $\mu=0$ case, and
that it should be a universal function depending only on the global
symmetries of the system.  We hope that our exact results could be an
important stepping stone to a deeper understanding of two-color QCD
(and maybe even real QCD) at nonzero $\mu$.

Recently, spectral sum rules for inverse Dirac eigenvalues have been
derived in the CFL phase of three-color QCD by two of us
\cite{Yamamoto2009}, but due to the sign problem it is not
straightforward to test them directly on the lattice. On the other
hand, the new sum rules obtained in this paper for two-color QCD can
in principle be tested on the lattice.  The lattice spacing has to be
sufficiently small compared to $1/\mu$ to avoid lattice artifacts, but
this seems to be mainly a technical problem and not a fundamental
one.

This paper is organized as follows. In section~\ref{sec:pre} we give
an overview of the symmetry breaking patterns depending on the
presence of the chiral condensate, the diquark condensate, and/or the
chemical potential (see also the discussion in \cite{Kogut2000}).  In
section~\ref{sec:efftheory}, after reviewing the hierarchy of scales
near the Fermi surface, we construct the low-energy effective
Lagrangian for the Nambu-Goldstone (NG) bosons at large $\mu$ based on
the pattern of chiral symmetry breaking caused by the diquark
condensate.  We also derive a mass formula connecting the mass of the
NG bosons to the quark mass and the fermion gap.  In
section~\ref{sec:finite} we define the new $\epsilon$-regime and
exactly determine the quark mass dependence of the partition function.
We then derive spectral sum rules for the inverse eigenvalues of the
Dirac operator at large $\mu$.  We conclude in
section~\ref{sec:conclusion}.  In the appendix we give a detailed
microscopic derivation of the mass terms appearing in the low-energy
effective Lagrangian.

\section{Symmetry breaking patterns}
\label{sec:pre}

Note that in this paper we discuss the continuum theory.  Symmetries
and their breaking pattern on the lattice depend on the lattice
regularization adopted for the fermions and must be analyzed
separately \cite{Hands1999,Hands2000}.

We first define our notation, which is based on the notation used in
\cite{Kogut2000}.  The fermionic part of the Lagrangian in Euclidean
space reads $\bar\psi(\D(\mu)+\mathcal{M})\psi$ with the
$\mu$-dependent Dirac operator 
\begin{equation}
  \label{Dirac_operator}
  \D(\mu)=\gamma_\nu D_\nu+\gamma_0\mu
\end{equation}
and the mass term
\begin{equation}
  \mathcal{M}=\frac{1}{2}(1+\gamma_5)M+\frac{1}{2}(1-\gamma_5)M^\dagger\,.
\end{equation}
Here, $\psi$ is a short-hand notation for $N_f$ flavors of two-color
Dirac spinor fields transforming in the fundamental representation of
$\SU(2)_\text{color}$.  The $\gamma_\nu$ are hermitian
$\gamma$-matrices, with $\gamma_5=\gamma_0\gamma_1\gamma_2\gamma_3$.
The covariant derivative $D_\nu$ is an anti-hermitian operator so that
the eigenvalues of $\gamma_\nu D_\nu$ are purely imaginary. $M$ is the
$N_f\times N_f$ quark mass matrix, and for real and degenerate quark
masses the mass term simplifies to $\mathcal{M}=m\1_{N_f}$.  

Because of the pseudo-reality of $\SU(2)$ we have $\D(\mu)\tau_2 C
\gamma_5=\tau_2 C \gamma_5 \D(\mu)^{*}$, where $C$ is the charge
conjugation operator and $\tau_2$ is a generator of
$\SU(2)_\text{color}$.  Together with chiral symmetry, $\{\gamma_5,
\D(\mu)\}=0$, this can be used to show that if $\lambda$ is one of the
eigenvalues of $\D(\mu)$, so are $-\lambda,\,\lambda^*$, and
$-\lambda^*$.  While this observation is sufficient to see that the
fermion determinant $\det(\D(\mu)+\mathcal M)$ is real, it does not
necessarily imply that two-color QCD is free of the sign problem; the
point is that $\lambda$ and $\lambda^*$ are degenerate if $\lambda$ is
real \cite{Hands2000}.  Consequently, a sign problem can arise in
two-color QCD with an odd number of flavors.  We also note that in the
simultaneous presence of a quark chemical potential $\mu$ and an
isospin chemical potential $\mu_{I}$, a sign problem can arise in
two-color QCD with an arbitrary number of flavors
\cite{Splittorff2001}.  Therefore, in this paper we only consider an
even number of flavors and $\mu_{I}=0$ so that the sign
problem is absent.

For $M=0$ and $\mu=0$, the fermionic part of the Lagrangian is
symmetric under $\U(2N_f)$. For $\mu\ne0$, this symmetry is broken
explicitly to $\U(N_f)_L\times\U(N_f)_R$.  In general, the chiral
symmetry of the Lagrangian is not realized in the ground state of the
theory.  Below we list the symmetries realized in various phases of
two-color QCD with $N_f$ flavors of Dirac fermions in the fundamental
representation
\cite{Peskin:1980gc,Kogut1999,Kogut2000}.\footnote{There is no
  $\SU(2)$ global anomaly \cite{Witten1982} since we deal with Dirac
  fermions.}\footnote{The
  symplectic group $\Sp(2n)=\{g\in U(2n)\:|\: g^TIg=I\}$, where $I$ is 
  defined in (\ref{I}). The dimension of $\Sp(2n)$ is $2n^2+n$, and its 
  center is $\Z_2$ irrespective of $n$.}
\begin{enumerate}\label{S_B_P}
\item $\mu=0$, $\langle\bar\psi\psi\rangle=0$,
  $\langle\psi\psi\rangle=0$: $\quad \SU(2N_f)\ [\supset
  (\Z_{2N_f})_A]$
\item $\mu=0$, $\langle\bar\psi\psi\rangle\ne0$,
  $\langle\psi\psi\rangle=0$: $\quad \Sp(2N_f)\ [\supset (\Z_2)_A]$
\item $\mu\ne0$, $\langle\bar\psi\psi\rangle=0$,
  $\langle\psi\psi\rangle=0$: $\quad \SU(N_f)_L\times \SU(N_f)_R\times
  \U(1)_B\ [\supset (\Z_{2N_f})_A]$
\item $\mu\ne0$, $\langle\bar\psi\psi\rangle\ne0$,
  $\langle\psi\psi\rangle=0$: $\quad \SU(N_f)_V\times \U(1)_B\
  [\supset (\Z_{2})_A]$
\item $\mu\ne0$, $\langle\bar\psi\psi\rangle=0$,
  $\langle\psi\psi\rangle\ne0$: $\quad \Sp(N_f)_L\times \Sp(N_f)_R\
  [\supset (\Z_2)_L \times (\Z_2)_R]$
\item $\mu\ne0$, $\langle\bar\psi\psi\rangle\ne0$,
  $\langle\psi\psi\rangle\ne0$: $\quad \Sp(N_f)_V\ [\supset (\Z_2)_B]$
\end{enumerate}
In the first two lines, $\U(1)_B$ is contained in $\SU(2N_f)$ and
$\Sp(2N_f)$, respectively.  $(\Z_{2N_f})_A$ is the anomaly-free
subgroup of the axial $\U(1)_A$ symmetry.  $(\Z_{2N_f})_A$ is already
contained both in $(\Z_{N_f})_L \times \U(1)_B$ and in $(\Z_{N_f})_R
\times \U(1)_B$.  Our notation is such that $\Sp(2)\sim\SU(2)$, therefore
$\Sp(n)$ is defined only if $n$ is even.  In the last two lines, an
even number of flavors is assumed. The fifth line is the case relevant 
for this paper, and the corresponding symmetry breaking pattern will be
discussed in more detail in section~\ref{sec:efftheory}.

The color-singlet diquark field $\psi\psi$ is a short-hand notation
for
\begin{equation}
  \psi\psi\equiv\epsilon_{ab}(\psi_a^T)^i C\gamma_5 I^{ij}\psi_b^j\,,
  \label{diquark}
\end{equation}
where $C$ is the charge-conjugation matrix, $a,b$ are color indices,
and $i,j$ are flavor indices, respectively.  The $N_f\times N_f$
symplectic matrix $I$ is defined as
\begin{equation}
  I=\begin{pmatrix}0&-\1\\\1&0\end{pmatrix},
  \label{I}
\end{equation}
where $\1$ is the $(N_f/2)\times(N_f/2)$ unit matrix. (For $N_f=2$, $I$ is equal to
$-i\sigma_2$.)  As long as $s$-wave condensation is assumed, there may
in principle be not only a color- and flavor-antisymmetric condensate
but also a color- and flavor-symmetric condensate.  However, the
single-gluon exchange interaction indicates that the antisymmetric 
condensate is energetically favored \cite{Ratti2004}, and thus we
will not consider the symmetric one in the following.

Besides the scalar diquark condensate $\psi^T C\gamma_5\psi$ in
(\ref{diquark}), there is in principle also a pseudo-scalar diquark
condensate $\psi^T C\psi$.  Although the two are not distinguished by
a single-gluon exchange interaction, it has been shown
\cite{Alford1998,Rapp1998} that the instanton-induced interaction
favors the former. Based on this observation we also neglect the
latter condensate in the following.\footnote{In \cite{Kogut1999} it
  was rigorously shown by QCD inequalities, which hold even at nonzero
  chemical potential owing to the positivity of the Euclidean
  path-integral measure of two-color QCD, that for massless quarks the
  lightest meson is the $0^+$ diquark, and thus a condensation (if it
  occurs) must occur in the channel $\psi^TC\gamma_5\psi$.}

\section{Effective theory for dense two-color QCD}
\label{sec:efftheory}

We now proceed to construct the low-energy effective Lagrangian for
two-color QCD at large $\mu$.  A similar construction for the CFL
phase of three-color QCD has been performed in
\cite{Casalbuoni1999e,Son2000,Son2000a,Schafer2002}.

\subsection{Hierarchy of scales}
\label{ahahaha}

We first give a summary of the energy spectrum near the Fermi surface.
For simplicity, the discussion in this subsection will mainly be restricted
to the chiral limit (i.e., $M=0$).  According to perturbative
calculations at large $\mu$ \cite{Son1999a,Schafer:1999fe} we have
$0\simeq \langle\bar\psi\psi\rangle\ll
\langle\psi\psi\rangle$.\footnote{$\langle\bar\psi\psi\rangle\simeq 0$
  was observed in lattice simulations even at intermediate $\mu$
  \cite{Kogut2001}.}  Therefore chiral symmetry is broken by the
\emph{diquark condensate}, not by the chiral condensate.  The symmetry
breaking pattern is (see section~\ref{sec:pre})
\begin{equation}
  \label{sym}
  \SU(N_f)_L\times \SU(N_f)_R\times \U(1)_B\times \U(1)_A \rightarrow
  \Sp(N_f)_L\times \Sp(N_f)_R\,. 
\end{equation}
From the seminal renormalization group analysis near the Fermi surface
\cite{Son1999a}, the fermion gap $\Delta$ for $\SU(2)_\text{color}$ is
found to be \cite{Schafer:1999fe}
\begin{equation}
  \Delta \sim \mu g^{-5}\exp\left(-\frac{2\pi^2}{g} \right),
  \label{estimate_of_gap}
\end{equation}
where $g\equiv g(\mu)$ is the small running coupling obtained from the
one-loop beta-function as
\begin{align}
  g(\mu)^2 \simeq \frac{12\pi^2}{(11-N_f)\ln(\mu/\Lambda_{\SU(2)})}
  \quad \text{for } \mu \gg \Lambda_{\SU(2)}\,.
\end{align}
We assume $N_f<11$ to ensure asymptotic freedom.  In particular, we
have
\begin{equation}
  \Lambda_{\SU(2)}\ll\Delta\ll\mu
\end{equation}
for sufficiently large $\mu$.\label{factor_c}

We first discuss the breaking of the $\U(1)_B$ and $\U(1)_A$
symmetries.  The $\U(1)_B$ symmetry is spontaneously broken, with an
associated NG boson $H$ that is gapless irrespective of the current
quark mass.  The $\U(1)_A$ symmetry is broken explicitly by the
anomaly/instantons for small $\mu$, but this explicit symmetry
breaking disappears as $\mu\to \infty$ because instantons are screened
in this limit \cite{Shuryak1982a}.  In the latter case the $\U(1)_A$
symmetry is broken spontaneously by the diquark
condensate, and we have an associated NG boson $\eta'$.  At
$\mu<\infty$ the $\eta'$ acquires a mass from the anomaly, but for
$\mu\gg\Lambda_{\SU(2)}$ the contribution of the anomaly to
$m_{\eta'}$ can be neglected,
and the $\eta'$ can be treated as a pseudo-NG boson
\cite{Son2000}.  In other words, at large $\mu$ it is a good
approximation to treat $\U(1)_A$ as a non-anomalous symmetry and
assume that the partition function is independent of the vacuum angle
$\theta$.

We now turn to the breaking of $\SU(N_f)_L\times \SU(N_f)_R$ to
$\Sp(N_f)_L\times \Sp(N_f)_R$.  The case of $N_f=2$ is special since
$\SU(2)\sim\Sp(2)$.  In this case there is no additional NG boson.  For
$N_f=4$ and larger, the spectrum contains additional NG bosons $\pi$
whose masses are lifted by a nonzero current quark mass.  Their total
number is
\begin{equation}
  2(N_f^2-1)-2(N_f^2/2+N_f/2)=N_f^2-N_f-2\,.
\end{equation}

Are there other light particles? Quarks acquire a gap $\Delta$ and
decouple from the dynamics near the Fermi surface. What about $\SU(2)$
gluons?  In two-color QCD, the diquark condensate (\ref{diquark}) is a
color singlet, so the $\SU(2)$ gluons do not acquire a mass through
the Higgs mechanism.  The absence below $\Delta$ of particles charged
under $\SU(2)$ indicates that the medium is transparent for $\SU(2)$
gluons and that neither Debye screening nor Meissner effect occur
\cite{Rischke2000}.  For these reasons, the low-energy dynamics of
gluons is simply described by the Lagrangian of $\SU(2)$ Yang-Mills
theory.  It has been found in \cite{Rischke2001}
that the confinement scale of in-medium $\SU(2)$ gluodynamics is
considerably diminished from the value $\sim\Lambda_{\SU(2)}$ at
$\mu=0$ to
\begin{equation}
  \Lambda'_{\SU(2)}\sim \Delta\exp \left(-\frac{2\sqrt{\mathstrut
        2}\pi}{11}\frac{\mu}{g\Delta} \right) \ll\Lambda_{\SU(2)}
\end{equation}
due to the polarization effect of the diquark condensate.  Since
$\mu/g\Delta$ is large, see (\ref{estimate_of_gap}), it follows that 
gluons become almost gapless at asymptotically large $\mu$.  A
more quantitative analysis \cite{Schafer2003} even suggests that 
gluons are lighter than the $\eta'$ for sufficiently large $\mu$.
However, since their coupling to NG bosons does not show up at 
leading order of the QCD weak-coupling calculations
\cite{Schafer2002,Schafer2003},  we assume in the following that
gluons do not interact with NG bosons.

The discussion of this section is summarized in table~\ref{table}.
The explanation of why $m_\pi^2=0$ or $\propto m^2$ for large $\mu$
will be given in section~\ref{sec:lagrangian}.

\TABLE{
  \begin{tabular}{|c||c|c|}
    \hline
    &$\mu\ll\Lambda_{\SU(2)}$&$\mu\gg\Lambda_{\SU(2)}$\\\hline
    chiral symmetry&spontaneously broken&spontaneously broken\\
    $m_\pi^2$&$\propto m$&$0\text{ or}\propto m^2$\\
    instantons&abundant&suppressed\\
    $\U(1)_B$ symmetry&intact&spontaneously broken\\
    $\U(1)_A$ symmetry&anomalous&spontaneously broken\\
    $\eta'$&heavy&light\\
    condensate&$0\simeq\langle\psi\psi\rangle\ll |\langle\bar\psi\psi\rangle|$&
    $0\simeq\langle\bar\psi\psi\rangle\ll\langle\psi\psi\rangle$\\
    mass gap\ (quarks)&$\sim\Lambda_{\SU(2)}$&$\sim\Delta\gg\Lambda_{\SU(2)}$\\
    mass gap\ (gluons)&$\sim\Lambda_{\SU(2)}$&$\sim\Lambda_{\SU(2)}'\ll\Lambda_{\SU(2)}$\\\hline
  \end{tabular}
  \caption{Qualitative differences in the physics of two-color QCD at
    large and small $\mu$.  $m$ denotes the (degenerate) current quark mass. $N_f$
    is assumed to be even.  
    See text for further explanations.
	In the second column, $\langle\psi\psi\rangle\ll
    |\langle\bar\psi\psi\rangle|$ only if $\mu \ll m_\pi/2$.}
  \label{table}
}

\subsection{Chiral Lagrangian}
\label{sec:lagrangian}

We now turn to the construction of the low-energy effective Lagrangian
associated with the symmetry breaking pattern \eqref{sym}.  We
introduce dimensionless color-singlet $N_f\times N_f$ matrix fields
$D_L$ and $D_R$, where $N_f$ is even and $>2$ (the case $N_f=2$ will
be considered separately at the end of this subsection).  In terms of the
microscopic fields, they can be expressed as
\begin{equation}
  (D_L)^{ij}\sim(\psi^T_L)^i C \psi_L^j\,, \qquad
  (D_R)^{ij}\sim(\psi^T_R)^i C \psi_R^j\,.
\end{equation}
Under $\SU(N_f)_{L}\times \SU(N_f)_{R}\times \U(1)_A\times \U(1)_B$
the quarks transform as
\begin{equation}
  \psi_L\to \ee^{i(\alpha+\beta)}g_L\psi_L\,,\qquad \psi_R\to
  \ee^{-i(\alpha-\beta)}g_R\psi_R\,,
\end{equation}
where $g_i\in\SU(N_f)_i$ ($i=L,R$) and the phases $\alpha$ and $\beta$
are associated with the $\U(1)_A$ and $\U(1)_B$ rotations,
respectively.  Thus
\begin{equation}
  D_L\to g_LD_Lg_L^T\ee^{2i(\alpha+\beta)}\,,\qquad 
  D_R\to g_RD_Rg_R^T\ee^{-2i(\alpha-\beta)}\,.
\end{equation}
It is convenient to split the fields into $\U(1)$ parts $A$, $V$ and
the rest,
\begin{equation}
  D_L\equiv\Sigma_LA^\dagger V\,,\qquad D_R\equiv\Sigma_R AV\,,
\end{equation}
so that
\begin{equation}
  \Sigma_i\to g_i\Sigma_i g_i^T\quad (i=L,R)\,,\qquad
  A\to A\ee^{-2i\alpha}\,,\qquad V\to V\ee^{2i\beta}\,.
  \label{trans-1}
\end{equation}
The mass term $\bar\psi_LM\psi_R+\bar\psi_RM^\dagger\psi_L$ is
invariant if we treat the mass matrix $M$ as a spurion field and let
\begin{equation}
  M\to g_LMg_R^\dagger \ee^{2i\alpha}\,.
  \label{trans-2}
\end{equation}
The effective Lagrangian composed of $\Sigma_L$, $\Sigma_R$, $A$, $V$,
and $M$ should be invariant under (\ref{trans-1}) and (\ref{trans-2}).
There is no invariant combination that contains an odd number of
factors of $M$, in contrast to \cite{Kogut2000} where the effective
Lagrangian contained a term linear in $M$.
This difference can be understood as follows.  As $N_f$ is even, we
can take $g_L=-1$ and $g_R=1$ (or $g_L=1$ and $g_R=-1$), under which
$M$ transforms as $M\to -M$ while $\Sigma_L$, $\Sigma_R$, $A$, and $V$
remain unchanged.  Therefore $M$ must appear in even powers.  The
point is that the chiral condensate $\langle\bar\psi_L\psi_R\rangle$
is negligible at large $\mu$; otherwise a NG field
$\tilde\Sigma\sim\bar\psi_L\psi_R$ would appear that transforms under
$g_L=-1$ as $\tilde\Sigma\to -\tilde\Sigma$ so that odd powers of $M$
could appear in the Lagrangian in combination with $\tilde\Sigma$.
(Strictly speaking, the assumption that $N_f$ is even is not essential
here.  We can reach the same conclusion from the fact that the
explicit breaking of $\U(1)_A$ by instantons vanishes at
large $\mu$, as discussed in \cite{Son2000}.)

At $O(M^2)$ the real-valued invariant combination is uniquely found to
be
\begin{equation}
  A^2\tr(M\Sigma_R M^T\Sigma_L^\dagger)+\text{c.c.}
  \label{mass_term}
\end{equation}
No mass term appears for $V$ reflecting the fact that $\U(1)_B$ is not
violated by a nonzero quark mass.  $\Sigma_L$ and $\Sigma_R$, which
are decoupled from each other in the chiral limit, are now coupled by $M$. 
Replacing $\Sigma_{L,R}$ by $\Sigma_{L,R}^T$ does not yield new invariants. 
This is because the diquark condensate is formed in a flavor
antisymmetric channel, i.e., $\Sigma^T_{L,R}=-\Sigma_{L,R}$ as already
noted in section~\ref{sec:pre}. 
In addition, we have $\Sigma\Sigma^\dagger=-\Sigma\Sigma^*=\1_{N_f}$ and
$\det\Sigma=1$ for $\Sigma\in \SU(N_f)/\Sp(N_f)$, which also greatly
reduces the number of nontrivial invariants.

We conclude that the effective Lagrangian in Minkowski space-time, to lowest order in the
derivatives and the quark masses, and assuming $N_f>2$ and even, is given
by\footnote{The Wess-Zumino-Witten term is necessary for completeness
  of the theory, but it is irrelevant to the ensuing analysis and  
  neglected in the following.}
\begin{align}
  \label{L}
  \mathcal{L}=\ &\frac{f_{H}^2}{2}\Big\{|\del_0 V|^2-v_{H}^2|\del_i
  V|^2\Big\} + \frac{N_ff_{\eta'}^2}{2}\Big\{|\del_0
  A|^2-v_{\eta'}^2|\del_i A|^2\Big\} \\
  &+\frac{f_{\pi}^2}{2}\tr\Big\{|\del_0\Sigma_L|^2-v_{\pi}^2|
  \del_i\Sigma_L|^2+(L\leftrightarrow R)\Big\} -c \Delta^2
  \Big\{A^2\tr(M\Sigma_R M^T\Sigma_L^\dagger)+\text{c.c.}\Big\}\,.\notag
\end{align}
Here, $f_{H}$, $f_{\eta'}$, and $f_{\pi}$ are the decay constants of
$H$, $\eta'$, and $\pi$, respectively, and the $v$'s are the
corresponding velocities originating from the absence of Lorentz
invariance in the medium.  The coefficient $c$ is 
calculable using the method of \cite{Schafer2002} and found to be
\begin{equation}
  c=\frac{3}{4\pi^2}\,.
  \label{c}
\end{equation}
The detailed derivation of (\ref{c}) is given in appendix~\ref{app_A}.
Note that $\mathcal{L}$ has no dependence on the $\theta$-parameter.
The effective chemical potential of $O(M^2)$ (the so-called
Bedaque-Sch\"afer term in the CFL phase of three-color QCD
\cite{Bedaque2002}) is not displayed here since it is suppressed by
$\sim 1/\mu$ at large $\mu$. Also, its contribution to the
finite-volume partition function starts at $O(M^4)$, which is
sufficiently small compared to the leading $O(M^2)$ term (see
section~\ref{sec:finite}).

From (\ref{L}) we can derive a mass formula for $\pi$ in the
flavor-symmetric case $M=m {\bf 1}_{N_f}$.  For this purpose, we define the 
NG fields $\pi_{i}^a$ ($i=L,R$, $a=1,\ldots,(N_f^2-N_f)/2-1$)
corresponding to the coset space $\SU(N_f)_{i}/\Sp(N_f)_{i}$
as\footnote{For $N_f=2$, $UIU^T=(\det U)I=I$ is constant, in
  accordance with the isomorphism $\SU(2)\sim\Sp(2)$.}
\begin{align}
  \Sigma_{i}=U_i I U_i^T\,, \qquad
  U_i=\exp\left(\frac{i \pi_i^a X^a}{2\sqrt{N_f} f_{\pi}} \right),
  \label{eq:pion-field}
\end{align}
where $I$ is defined in (\ref{I}) and the $X^a$ are the generators of
the coset $\SU(N_f)/\Sp(N_f)$ satisfying the normalization $\tr(X^a
X^b)=N_f\delta^{ab}$ and the relation $X^a I = I (X^a)^T$.  Since the
mass term in (\ref{L}) mixes $\pi_L$ and $\pi_R$, it is necessary to
diagonalize the mass matrix for $\pi_{L,R}^a$ to obtain the genuine
mass eigenvalues by setting
\begin{equation}
  \Pi^a = \frac{1}{\sqrt{2}}(\pi_L^a + \pi_R^a)\,, \qquad 
  \tilde \Pi^a = \frac{1}{\sqrt{2}}(\pi_L^a - \pi_R^a)\,.
\end{equation}
The resulting mass formula for $\Pi^a$ and $\tilde \Pi^a$ 
reads
\begin{align}
  \label{eq:mass-formula}
  m_{\Pi^a}=0\,, \qquad
  f_{\pi}^2 m_{\tilde \Pi^a}^2 = 4c \Delta^2 m^2\,.
\end{align}
The massless modes $\Pi^a$ correspond to simultaneous rotations
of $\pi_L^a$ and $\pi_R^a$ in the same direction, while the massive
modes $\tilde \Pi^a$ correspond to rotations in the opposite direction (a
similar discussion can be found in \cite{Yamamoto2007a}).\footnote{If
  the quark masses are nondegenerate, the $\Pi^a$ modes acquire
  masses.} For the massive modes $\tilde \Pi^a$,
(\ref{eq:mass-formula}) is of exactly the same form as the expression
derived in \cite{Son2000}, except for the numerical factor of $c$.
However, our ``pions'' for two colors are two-quark ($q q$) states,
while those for three colors are four-quark ($\bar q \bar q q q$)
states in \cite{Son2000}.

Similarly, we can also derive a mass formula for the $\eta'$ boson
associated with the spontaneous breaking of $\U(1)_A$.
We again assume the flavor-symmetric case $M=m {\bf 1}_{N_f}$.
The $\eta'$ field is defined by
\begin{equation}
A=\exp\left(i \frac{\eta'}{\sqrt{N_f}f_{\eta'}} \right).
\end{equation}
Expanding to second order in terms of the $\eta'$ field in (\ref{L}) 
yields
\begin{equation}
  \label{eq:mass-formula2}
  f_{\eta'}^2 m_{\eta'}^2 = 4c \Delta^2 m^2\,. 
\end{equation}
Since $f_{\pi, \eta'} \sim \mu$ \cite{Son2000} we obtain from
(\ref{eq:mass-formula}) and (\ref{eq:mass-formula2}) the relation
\begin{equation}
  m_{\Pi, \eta'} \sim \frac{m\Delta}\mu\:.  
\end{equation}

Finally, we consider the simpler case $N_f=2$.  In this case, the
``pions'' disappear.  The first line of (\ref{L}) requires no change
while the second line is replaced by
\begin{equation}
  \label{c'}
  -c' \Delta^2 \Big\{(\det M)A^2+\mbox{c.c.}\Big\}\quad\text{with}\quad
  c'=\frac{3}{2\pi^2}\,.
\end{equation}
The above value of $c'$ corrects the value of $4/3\pi^2$ given in
\cite{Schafer2003}.\footnote{We thank Thomas Sch\"afer for a
  communication on this point.}

\section{Finite-volume analysis}
\label{sec:finite}

\subsection{Microscopic domain}

In this subsection we identify the microscopic domain of dense
two-color QCD and discuss some related intricacies.  In the following,
we neglect the $H$ boson and the gluonic sector since they decouple
from the dynamics of the NG bosons we are interested in and do not
contribute to the quark mass dependence of the partition function.

Let us study two-color QCD on a Euclidean four-dimensional torus $\beta\times
L\times L\times L$ with linear spatial extent $L$ and near-zero
temperature $\beta=1/T\sim L$.  As is well known, the chiral limit and
the thermodynamic limit do not commute in general.  A particularly
interesting regime to consider is the ``$\varepsilon$-regime'' (or
``microscopic domain'') of the theory in which the zero-momentum modes
of the NG bosons dominate the partition function \cite{Gasser:1987ah}.
In the present setting, this regime is defined by the inequalities
\begin{equation}
  \frac{1}{\Delta}\ll L\ll\frac{1}{m_{\Pi, \tilde \Pi, \eta'}}\,,
  \label{micro}
\end{equation}
where $m_{\Pi, \tilde \Pi, \eta'}$ 
is the mass of the NG bosons ($\Pi, \tilde \Pi, \eta'$) at large $\mu$.  
The first inequality, $1/\Delta\ll L$,
means that the contribution of heavy non-NG bosons to the partition
function can be neglected.  The second inequality means that if the
box size is much smaller than the Compton wave length of the NG
bosons, the functional integral over NG boson fields reduces to the
zero-mode integral over the coset space $\SU(N_f)_L/\Sp(N_f)_L\times
\SU(N_f)_R/\Sp(N_f)_R\times \U(1)_A$.

We briefly comment on the meaning of symmetry breaking assumed here.
It is well known that a global symmetry cannot be broken spontaneously
in a finite volume.\footnote{Although there are some exceptions such 
as large-$N$ transitions \cite{Gross:1980he} and dynamical supersymmetry 
breaking \cite{Witten:1982df}.}
However, in two-color QCD the diquark condensate is a color singlet,
which enables us to study diquark condensation phenomena by adding a
\emph{gauge-invariant} external symmetry-breaking field.  This method
is commonly used in lattice simulations, which are always restricted
to a finite volume.  This situation is in stark contrast to the CFL
phase of three-color QCD, where the four-quark state composed of left-
and right-handed diquark condensates is a color singlet, but the
diquark condensate itself is not \cite{Yamamoto2009}.

\subsection{Massless spectral sum rules}

In this subsection we determine the dependence of the finite-volume
partition function on the light quark masses and use it to derive a set
of exact spectral sum rules for the eigenvalues of the Dirac operator
$\D(\mu)$ defined in (\ref{Dirac_operator}).  Let us first consider
$N_f\geq 4$ (the case $N_f=2$ will be discussed at the end of this
subsection).  Since the symmetry breaking pattern $\SU(n)\to \Sp(n)$
is the same as at $\mu=0$ we can directly borrow calculational tools
from \cite{Smilga1995}.
The finite-volume partition function in the microscopic domain
\eqref{micro} is given by
\begin{align}
  Z(M) 
  & =\int\limits_{\U(1)_A}\hspace{-2mm}dA\hspace*{-5mm}
  \int\limits_{\hspace*{5mm}\SU(N_f)_L/\Sp(N_f)_L}\hspace{-13mm}
  d\Sigma_L\hspace{3mm}
  \int\limits_{\hspace*{5mm}\SU(N_f)_R/\Sp(N_f)_R}\hspace{-13mm} 
  d\Sigma_R\hspace{3mm}
  \exp\left[V_4c \Delta^2 \Big\{A^2\tr(M\Sigma_R
    M^T\Sigma_L^\dagger)+\text{c.c.}\Big\}\right] \notag \\
  &=\int\limits_{\U(1)_A}\hspace{-2mm}dA
  \int\limits_{\SU(N_f)_L}\hspace{-4mm} dU_L
  \int\limits_{\SU(N_f)_R}\hspace{-4mm} dU_R \hspace{5pt}
  \exp \left[-V_4c \Delta^2
    \Big\{A^2\tr(MU_RIU^T_RM^TU_L^*IU_L^\dagger)
    +\text{c.c.}\Big\}\right],
  \label{Z(M)-}
\end{align}
where $Z$ is normalized to 1 in the chiral limit (i.e., all group
volumes are defined to be unity) and $V_4\equiv L^3\beta$ is the
Euclidean space-time volume.  Note that $Z$ has no $\theta$-dependence.
In going from the first to the second line of \eqref{Z(M)-}, 
we have used $\Sigma=UIU^T$ as in \eqref{eq:pion-field}, but for
computational convenience we have chosen $U\in\SU(N_f)$ rather than
$U\in\SU(N_f)/\Sp(N_f)$.  The integral over $U$ is independent of the
additional degrees of freedom we have introduced because $UIU^T=I$ for
$U\in\Sp(N_f)$.

From the invariance property of the Haar measure, (\ref{Z(M)-}) can be
expanded as
\begin{equation}
  Z(M)=1+c_2\,\tr(M^\dagger M)+c_4^{(1)}(\tr M^\dagger
  M)^2+c_4^{(2)}\tr(M^\dagger M)^2 +O\big(M^6\big)\,.
\end{equation}
However, all terms of $O(M^{4k-2})$ ($k=1, 2, \ldots $) vanish because
of the $\U(1)_A$ integral.  In particular $c_2=0$, so the first
nontrivial coefficients are $c_4^{(1)}$ and $c_4^{(2)}$, both
proportional to $(V_4c\Delta^2)^2$.  To figure out their
proportionality constants, we use the formulas given in
\cite[(4.11)-(4.17)]{Smilga1995},
\begin{equation}
  \int\limits_{U(2N)}\hspace{-2mm}dU\, \exp\big\{\tr(YUIU^T)
  +\text{c.c.}\big\}=1+\frac{2}{2N-1}\tr(Y^\dagger Y)+O(Y^4)
  \label{borrowed}
\end{equation}
for any $2N\times 2N$ antisymmetric matrix $Y$. ($I$ is given in
(\ref{I}) but here has dimension $2N$.)  Comparing (\ref{Z(M)-}) and
(\ref{borrowed}) suggests the substitutions $U\to AU_R$, $N\to N_f/2$,
and $Y\to -V_4c \Delta^2 M^TU_L^*IU_L^\dagger M$ ($Y$ is
antisymmetric).  We find
\begin{align}
  Z(M)&=\int\limits_{\SU(N_f)_L}\hspace{-4mm}dU_L
  \,\left\{1-(V_4c\Delta^2)^2\frac{2}{N_f-1}\tr(M^\dagger U_LIU_L^TM^*
    \cdot M^TU_L^*IU_L^\dagger M)\right\}+O(M^8)
  \nonumber \\
  &=1+(V_4c\Delta^2)^2\frac{2}{(N_f-1)^2}\Big\{(\tr M^\dagger
  M)^2-\tr(M^\dagger M)^2\Big\}+O(M^8)\,.
  \label{Z_expansion}
\end{align}
We note in passing that the above combination $(\tr M^\dagger
M)^2-\tr(M^\dagger M)^2$ also appeared in our previous work for the
CFL phase \cite{Yamamoto2009}.  The above expansion is to be matched
later against the one derived directly from the fundamental QCD
Lagrangian.

As a special case, let us consider the case of a flavor-symmetric mass
term $M=m\1_{N_f}$ with real $m$. Now one of the two $\SU(N_f)$
integrals (which corresponds to the integral over the massless modes
$\Pi^a$ in (\ref{eq:mass-formula})) trivially drops out,
leaving the far simpler expression
\begin{equation}
  Z(m)=\int\limits_{U(N_f)}\hspace{-2mm}dU\,\exp \left[-V_4c \Delta^2
    m^2\Big\{\tr(UIU^TI)+\text{c.c.}\Big\}\right].
  \label{equal_mass}
\end{equation}
A small consistency check can be done using (\ref{borrowed}) in
(\ref{equal_mass}) to obtain (\ref{Z_expansion}) with $M=m\1_{N_f}$.  This expression is
of the same form as the one at $\mu=0$ \cite[(4.11)]{Smilga1995},
which can be rewritten as a Pfaffian \cite[(5.13)]{Smilga1995}.  By
the same token (with the replacements $x \rightarrow
4c\Delta^2V_4m^2$, $\nu \rightarrow 0$, and $2N_f \rightarrow N_f$,
respectively, in \cite[(5.13)]{Smilga1995}) we readily obtain the simple
expression\footnote{The Pfaffian of an $n\times n$ antisymmetric
  matrix $X$ (for even $n$) is defined as
\begin{equation}
  \pf(X)=\frac{1}{2^{n/2}(n/2)!}\sum_{\sigma}
  \sgn(\sigma)X_{\sigma(1)\sigma(2)}\dots X_{\sigma(n-1)\sigma(n)}\,,
\end{equation}
where the sum runs over all permutations $\sigma$ of
$1,\,2,\,\dots,n$.  Note that this definition differs by a factor of
$1/2^{n/2}$ from \cite[(A.8)]{Smilga1995}.}
\begin{equation}
\label{eq:pf}
  Z(m)=\frac{1}{(N_f-1)!!}\pf(A)\,, 
\end{equation}
where $A$ is an $N_f\times N_f$ antisymmetric matrix with entries
\begin{equation}
  A_{pq}\equiv (q-p)I_{p+q}(4c\Delta^2V_4m^2)\,,\qquad
  p,q=-\frac{N_f-1}{2},\ldots, \frac{N_f-3}{2},\frac{N_f-1}{2}
  \label{A_elements}
\end{equation}
and $I_{p+q}$ denotes a modified Bessel function.  
The analytical calculation of (\ref{Z(M)-}) for non-degenerate quark masses 
(as has been done elegantly at $\mu=0$ for Dyson index $\beta_D=2$
\cite{Brower1981b,Jackson1996a,Verbaarschot2007,Akuzawa1998} as well
as for $\beta_D=1,\,4$ \cite{Nagao2000a}) is left for future work.

Having determined the light quark mass dependence of the partition
function, we can now derive spectral sum rules for two-color QCD at
large $\mu$. The philosophy is well-known: By adding a weak external
field to a given physical system and measuring its response thereto,
one can gain information about properties of the system \emph{in the
  absence} of the external field.  In the present situation, the
external field is a non-zero quark mass, and we can learn something
about the properties of the Dirac spectrum \emph{in the chiral
  limit}.

We start with the properly normalized partition function expressed in
terms of the microscopic degrees of freedom,
\begin{equation}
  Z(M)=\int [\mathcal{D}A]\det \Big[\D(\mu)
  +\frac{1+\gamma_5}{2}M+\frac{1-\gamma_5}{2}M^\dagger\Big]\,
  \ee^{-S_g}\bigg/ \int [\mathcal{D}A]\,{\det}^{N_f}\D(\mu)\, \ee^{-S_g}\,,
\end{equation}
where $S_g\equiv\int d^4x\,F^a_{\mu\nu}F^a_{\mu\nu}/4$ is the gluonic
action.  Let us denote the complex eigenvalues of $\D(\mu)$ by
$i\lambda_n$, where the $\lambda_n$ are real for $\mu=0$.  In
accordance with the discussion in section~\ref{ahahaha}, we assume
that at large $\mu$ there are no exact zero modes, i.e., all
$\lambda_n$ are nonzero.  Since all eigenvalues come in pairs
$(i\lambda_n,-i\lambda_n)$, we have
\begin{align}
  Z(M)&=\int [\mathcal{D}A]\ {\prod_{n}}'\det(\lambda_n^2+M^\dagger
  M)\ \ee^{-S_g}\bigg/ \int [\mathcal{D}A]\
  \Big({\prod_{n}}'\lambda_n^2\Big)^{N_f}\ \ee^{-S_g} \notag \\
  &=\biggl\langle{\prod_{n}}'\det\left(1+\frac{M^\dagger
        M}{\lambda_n^2}\right)\biggr\rangle\,,
  \label{eq:Z}
\end{align}
where $\prod'_n$ (and later $\sum'_n$) denotes the product (sum) over
all eigenvalues $\lambda_n$ with $\re\lambda_n>0$, and
\begin{equation}
  \label{eq:av}
  \langle\mathcal O\rangle\equiv\int
  [\mathcal{D}A]\,\mathcal{O}\,\Big({\prod_n}'\lambda_n^2\Big)^{N_f}
  \,\ee^{-S_g} \bigg/
  \int[\mathcal{D}A]\,\Big({\prod_n}'\lambda_n^2\Big)^{N_f}\ee^{-S_g}\,.
\end{equation} 
Because of the pseudo-reality of $\SU(2)$, the measure in \eqref{eq:av}
is real and positive definite (for even $N_f$).  Using the relation
\begin{equation} 
  \det(1+\epsilon)=1+\tr\epsilon+\frac{1}{2}
  [(\tr \epsilon)^2-\tr\epsilon^2] + O(\epsilon^3)\,,
\end{equation}
$Z(M)$ in (\ref{eq:Z}) can be expanded in terms of the quark mass
matrix $M$ as
\begin{align}
  Z(M)&=1+\biggl\langle{\sum_{n}}'\frac{1}{\lambda_n^2}\biggr\rangle
  \tr M^\dagger M
  +\biggl\langle{\sum_{m<n}}'\frac{1}{\lambda_m^2\lambda_n^2}
  \biggr\rangle(\tr M^\dagger M)^2 \notag \\
  &\quad +\frac12\biggl\langle{\sum_{n}}'\frac{1}{\lambda_n^4}\biggr\rangle
  \Big\{(\tr M^\dagger M)^2-\tr (M^\dagger M)^2\Big\}+O(M^6)\,.
  \label{Z_expansion'}
\end{align}
Comparing the formal expansion (\ref{Z_expansion'}) with
(\ref{Z_expansion}), we obtain novel spectral sum rules for two-color
QCD at large $\mu$,\footnote{Sum rules for complex Dirac eigenvalues
  have been obtained earlier in a different context, i.e., at small
  $\mu$ in the conventional microscopic domain $m_\pi,\mu\ll
  1/L\ll\Lambda_\text{QCD}$, see, e.g., \cite{Akemann2007}.}
\begin{gather}
  \biggl\langle{\sum_{n}}'\frac{1}{\lambda_n^2}\biggr\rangle
  =\biggl\langle{\sum_{m<n}}'\frac{1}{\lambda_m^2\lambda_n^2}\biggr\rangle
  =\biggl\langle{\sum_{n}}'\frac{1}{\lambda_n^6}\biggr\rangle=0\,,
  \label{sum_1} \\
  \biggl\langle{\sum_{n}}'\frac{1}{\lambda_n^4}\biggr\rangle=(4c
  \Delta^2V_4)^2\frac{1}{4(N_f-1)^2}\,,
  \label{sum_2}
\end{gather}
which can in principle be continued to arbitrary orders of $M$.  Note
again that we have assumed even $N_f$ with $N_f\ge4$.  The sums
appearing in (\ref{sum_1}) and (\ref{sum_2}) are in fact real-valued
for any \emph{fixed} gauge configuration, i.e., before averaging over
gauge fields.  This can be seen from the pseudo-reality of $\SU(2)$:
If $\lambda$ is one of the eigenvalues, so is $\lambda^*$.

Let us now define the spectral density
\begin{equation}
  \label{eq:rho}
  \rho(\lambda)=\biggl\langle\sum_n\delta^2(\lambda-\lambda_n)\biggr\rangle\,,
\end{equation}
which is positive definite since the path-integral measure is (and
thus $\rho$ allows for a natural probabilistic interpretation). 
Note that the gauge average in \eqref{eq:rho} involves 
massless quarks, see \eqref{eq:av}.  The case of nonzero masses will be 
addressed in section \ref{sec:massive}.

Then (\ref{sum_2}) becomes
\begin{equation}
  \frac{1}{(4c\Delta^2V_4)^2}\int_{\mathbb{C}_+}
  d^2\lambda\:\frac{\rho(\lambda)}{\lambda^4}
  =\int_{\mathbb{C}_+}d^2z\:\frac{\rho_s^{V_4}(z)}{z^4}
  =\frac{1}{4(N_f-1)^2}
  \label{rescaled_sum}
\end{equation}
with $\mathbb{C_+}\equiv\{w\in\mathbb{C}\:|\:\re w>0\}$ and
\begin{equation}
  \label{eq:rhos}
  \rho_s^{V_4}(z)\equiv\frac{\pi^2}{3\Delta^2V_4}\,
  \rho\biggl(\frac{\pi z}{\sqrt{3\Delta^2V_4}}\biggr)\,,
\end{equation}
where the numerical factor $4c=3/\pi^2$ has been included in the definition to
simplify some analytical results that will be derived in section
\ref{sec:massive}.  Equation~(\ref{rescaled_sum}) implies
the existence of a microscopic spectral density defined by
\begin{equation}
  \rho_s(z)\equiv\lim_{V_4\to\infty}\rho_s^{V_4}(z).
  \label{m_s_d}
\end{equation}
Since the sum rule (\ref{rescaled_sum}) is determined solely by the 
symmetry breaking pattern induced by the diquark condensate, we expect 
that (\ref{m_s_d}) is a \emph{universal} function fully determined by the global 
symmetries and independent of the details of the microscopic interactions. 
Our conjecture is corroborated by various facts known at $\mu=0$ \cite{Verbaarschot:2000dy}:
\begin{itemize}
\item The Leutwyler-Smilga sum rules are satisfied to good accuracy 
by the Dirac eigenvalues computed in the instanton liquid model, 
a phenomenological model of QCD.
\item The microscopic spectral correlation functions derived from chiral random matrix theory ($\chi$RMT) 
generate the Leutwyler-Smilga sum rules exactly.  Furthermore, they coincide with 
those derived from the finite-volume partition function of QCD to
leading order in the $\epsilon$-regime, 
because $\chi$RMT and QCD have the same global symmetries.
\item The microscopic spectral correlation functions predicted by
  $\chi$RMT have been verified in a number of lattice QCD simulations.
\end{itemize}
It is thus natural to expect that the way the thermodynamic limit of 
the spectral density near zero is approached is universal at large $\mu$, too. 
Further insights would be obtained if we could construct the
appropriate random matrix model, a topic we hope to address
elsewhere.

It is interesting to ask how the vanishing sum rules (\ref{sum_1}) can
be understood in terms of the symmetries of the distribution of the
eigenvalues $\lambda_n$. One might conjecture that the eigenvalues
occur not only in quartets $(\lambda,-\lambda,\lambda^*,-\lambda^*)$
but in octets $(\pm \lambda,\pm i\lambda,\pm \lambda^*,\pm
i\lambda^*)$.  This would be a sufficient condition for all averages
of the type $\lambda^{-4k+2}$ ($k=1, 2, \ldots$) to vanish.  However,
this idea is flawed, as can be understood from
\begin{equation}
  \biggl\langle\biggl({\sum_{n}}'\frac{1}{\lambda_n^2}\biggr)^2
  \biggr\rangle=(4c\Delta^2V_4)^2\frac{1}{4(N_f-1)^2}\,,
\end{equation}
which follows from (\ref{sum_1}) and (\ref{sum_2}). 
Thus the symmetries are not realized per configuration,
but only after averaging.  We therefore turn to the spectral density
$\rho(\lambda)$. It satisfies $\rho(\lambda)=\rho(\lambda^*)
=\rho(-\lambda)=\rho(-\lambda^*)$.  We now show that (\ref{sum_1})
follows if $\rho(\lambda)$ also satisfies
$\rho(\lambda)=\rho(i\lambda)$,
\begin{equation}
  \int d^2\lambda\:\frac{\rho(\lambda)}{\lambda^{4k-2}}
  =\int d^2\lambda\:\frac{\rho(i\lambda)}{(i\lambda)^{4k-2}}
  =-\int d^2\lambda\:\frac{\rho(\lambda)}{\lambda^{4k-2}}=0\,.
\end{equation}
The assumption $\rho(\lambda)=\rho(i\lambda)$ does not contradict
(\ref{sum_2}). We speculate that it is indeed a property of $\rho$.
In view of so many symmetries, it is tempting to assume that
$\rho(\lambda)=\rho(|\lambda|)$. Then
\begin{equation}
  \int d^2\lambda\:\frac{\rho(\lambda)}{\lambda^{4}}
  =\int{|\lambda|d|\lambda|}\:\frac{\rho(|\lambda|)}{|\lambda|^{4}}
  \oint d\varphi\,\ee^{-4i\varphi}=0\,,
\end{equation}
which contradicts (\ref{sum_2}).  Thus $\rho(\lambda)$ is anisotropic.

We finally consider $N_f=2$. According to (\ref{c'}), the microscopic
partition function (\ref{Z(M)-}) is given by
\begin{align}
  Z(M)&=\int\limits_{\U(1)_A}\hspace{-2mm}dA\, \exp\Big\{
  V_4c'\Delta^2\Big[(\det M)A^2+\mbox{c.c.}\Big]\Big\} \notag \\
  &=I_0(2V_4c' \Delta^2 |\det M|) \notag\\
  &=1+\frac{1}{\,2\,}(V_4c'\Delta^2)^2\Big\{(\tr M^\dagger M)^2
  -\tr(M^\dagger M)^2\Big\}+O(M^8)\,,
  \label{Z(M):N_f=2}
\end{align}
where we used the Cayley identity for $2\times 2$ matrices. Comparing
(\ref{Z(M):N_f=2}) with (\ref{Z_expansion'}) we are again led to the
vanishing sum rules (\ref{sum_1}) and to
\begin{equation}
  \left\langle{\sum_{n}}'\frac{1}{\lambda_n^4}\right\rangle
  =(c'\Delta^2V_4)^2\,,
\label{sum_Nf=2}
\end{equation}
which again hints at the existence of the universal microscopic
spectral density (\ref{m_s_d}).  Note that \eqref{sum_Nf=2} is
actually the same as \eqref{sum_2} with $N_f=2$ because $c'=2c$.  The
latter equality follows from $\tr(MIM^TI)=-2\det M$ for $N_f=2$, see
\eqref{eq:mass2} and \eqref{eq:mass3}.  In other words, the analysis
of this subsection for $N_f\ge4$ holds for $N_f=2$ as well if we set
$\Sigma_L=\Sigma_R=I$ (corresponding to the trivial coset space
$\SU(2)/\Sp(2)$) and $U_L=U_R=\1$ in \eqref{Z(M)-}, and $U_L=\1$ in
\eqref{Z_expansion}.  Of course, the physics is different since there
are no ``pions'' for $N_f=2$.

At first sight, the above results are counter-intuitive in the
following sense. Since the $i\lambda_n$ are the eigenvalues of the
Dirac operator $\gamma_\nu D_\nu+\gamma_0\mu$ for asymptotically large
$\mu$, one naively expects $\lambda_n\sim \mu$, and the inverse
moments of the eigenvalues should \emph{decrease} for larger $\mu$.
But our exact results imply exactly the opposite!\footnote{$\Delta$
diverges as $\mu\to\infty$, see (\ref{estimate_of_gap}).} --- What
we have overlooked is the existence of the Fermi surface.  The typical
momentum is comparable to the Fermi momentum, hence $\gamma_\nu D_\nu$
is not necessarily small compared to $\gamma_0\mu$.  Delicate
cancellations between these two terms lead to the above nontrivial sum
rules.

\subsection{Massive spectral sum rules}
\label{sec:massive}

It is known that the massless spectral sum rules at $\mu=0$ due to
Leutwyler and Smilga can be generalized to the massive case by an
appropriate rescaling of masses \cite{Shuryak1993}, and that the
double-microscopic spectral correlations (called so because masses and
eigenvalues are rescaled simultaneously as the volume is taken to
infinity) derived from random matrix theory reproduce these sum rules
\cite{Damgaard1998,Damgaard1998a,Wilke:1997gf}. 
Here we show that the new spectral
sum rules we derived at large $\mu$ can be generalized to the massive
case in a similar fashion. Such an extension will be useful in future
comparisons with lattice gauge simulations, where light dynamical 
fermions are computationally expensive.

We begin with the simplest case, i.e., $N_f=2$ with masses $m_1$ and
$m_2$. According to the result in the previous subsection, we have
\begin{align}
  Z&=I_0(\alpha m_1m_2) \quad\text{with}\quad
  \alpha=2c'\Delta^2V_4=\frac{3\Delta^2V_4}{\pi^2} \notag\\
  &=\frac{1}{\mathcal{N}}\int [\mathcal{D}A]
  \,{\prod_n}'\big[(\lambda_n^2+m_1^2)(\lambda_n^2+m_2^2)\big]\ee^{-S_g}\,,
  \label{massive_measure}
\end{align}
where $\mathcal{N}$ is a mass-independent normalization factor. Below
we represent expectation values with respect to the massive measure by
$\langle\!\langle\cdots\rangle\!\rangle$.

Differentiation of $Z$ by the masses yields massive spectral sum
rules, e.g.,\footnote{We repeatedly used the identities
  $I_n'(x)=[I_{n-1}(x)+I_{n+1}(x)]/2$ and
  $I_n(x)/x=[I_{n-1}(x)-I_{n+1}(x)]/2n$.}
\begin{align}
  \frac{\del}{\del(m_1^2)}\log
  Z&=\alangle{\sum_n}'\frac{1}{\lambda_n^2+m_1^2}\arangle
  =\frac{(\alpha m_2)^2}{4}\frac{I_0(\alpha m_1m_2)-I_2(\alpha
    m_1m_2)}{I_0(\alpha m_1m_2)}\,,
  \label{1..}
  \\
  \frac{1}{Z}\frac{\del^2 Z}{\del(m_1^2)\del(m_2^2)}&
  =\alangle{\sum_n}'\frac{1}{\lambda_n^2+m_1^2}{\sum_k}'\frac{1}{\lambda_k^2+m_2^2}\arangle
  =\frac{\alpha^2}{4}\,,
  \label{2..}
  \\
  \frac{1}{Z}\frac{\del^2 Z}{\del(m_1^2)^2}&
  =\alangle{\sum_{n\not=k}}'\frac{1}{(\lambda_n^2+m_1^2)(\lambda_k^2+m_1^2)}\arangle
  \notag\\
  &=\frac{(\alpha m_2)^4}{96}\frac{3I_0(\alpha m_1m_2)-4I_2(\alpha m_1m_2)+I_4(\alpha m_1m_2)}{I_0(\alpha m_1m_2)}\,.
  \label{3..}
\end{align}
For degenerate masses ($m_1=m_2\equiv m$) we have $Z=I_0(\alpha m^2)$
so that
\begin{equation}
  \frac{1}{Z}\frac{\del^2Z}{\del(m^2)^2}
  =\alangle-2{\sum_n}'\frac{1}{(\lambda_n^2+m^2)^2}+4\Big({\sum_n}'\frac{1}{\lambda_n^2+m^2}\Big)^2\arangle
  =\frac{\alpha^2}{2}\frac{I_0(\alpha m^2)+I_2(\alpha m^2)}{I_0(\alpha m^2)}\,.
  \label{4..}
\end{equation}
In the massless limit equations \eqref{1..}--\eqref{4..} are
consistent with the massless sum rules (\ref{sum_1}) and
(\ref{sum_Nf=2}).  In terms of rescaled dimensionless variables,
\begin{equation}
  \label{eq:rescaled}
  z_n\equiv\lambda_n\frac{\sqrt{3\Delta^2V_4}}\pi\,,\qquad 
  \m_i\equiv m_i\frac{\sqrt{3\Delta^2V_4}}\pi\,,
\end{equation}
equations \eqref{1..}--\eqref{4..} read
\begin{align}
  \alangle{\sum_n}'\frac{1}{z_n^2+\m_1^2}\arangle
  &=\frac{\m_2^2}4\,\frac{I_0(x)-I_2(x)}{I_0(x)}\quad\text{with}\quad
  x=\m_1\m_2\,,
  \label{1...}
  \\
  \alangle{\sum_n}'\frac{1}{z_n^2+\m_1^2}{\sum_k}'\frac{1}{z_k^2+\m_2^2}\arangle
  &=\frac14\,,
  \label{2...}
  \\
  \alangle{\sum_{n\not=k}}'\frac{1}{(z_n^2+\m_1^2)(z_k^2+\m_1^2)}\arangle
  &=\frac{\m_2^4}{96}\,
  \frac{3I_0(x)-4I_2(x)+I_4(x)}{I_0(x)}\,,
  \label{3...}
  \\
  \alangle-{\sum_n}'\frac{1}{(z_n^2+\m^2)^2}+2\Big({\sum_n}'&\frac{1}{z_n^2+\m^2}\Big)^2\arangle
  =\frac{I_0(\m^2)+I_2(\m^2)}{4I_0(\m^2)}\,.
  \label{4...}
\end{align}
It is surprising that the right-hand side of (\ref{2...}) is
independent of the masses.

We now define the double-microscopic spectral density by
\begin{equation}
  \label{eq:rhos2}
  \rho_s^{(N_f)}(z;\m_1,\dots,\m_{N_f})\equiv\lim_{V_4\to\infty}
  \frac{\pi^2}{3\Delta^2V_4}\rho\bigg(\frac{\pi z}{\sqrt{3\Delta^2V_4}}\bigg)\bigg|_{\m_i=m_i\frac{\sqrt{3\Delta^2V_4}}\pi\text{ fixed}}\,.
\end{equation}
As mentioned before, the rescaling is the same for $N_f=2$ and
$N_f\ge4$ because $c'=2c$, and thus \eqref{eq:rhos2} applies to all
(even) $N_f$.  From the above sum rules we can derive several
equalities characterizing $\rho_s^{(2)}$, e.g.,
\begin{equation}
  \int_{\mathbb{C}_+}d^2z\,\frac{\rho_s^{(2)}(z;\m,\m)}{(z^2+\m^2)^2}
  =\frac{I_0(\m^2)-I_2(\m^2)}{4I_0(\m^2)}\,,
\end{equation}
which is obtained from (\ref{4...}) and from (\ref{2...}) in the limit
$\m_1,\m_2\to\m$.  The determination of $\rho_s^{(N_f)}$ itself
requires further theoretical analysis and is left for future work.

For larger $N_f$ the explicit expressions become increasingly involved. For $N_f=4$ with equal masses, we have
\begin{equation}
  \alangle{\sum_n}'\frac{1}{z_n^2+\m^2}\arangle=\frac{2I_0(y)I_1(y)-3I_1(y)I_2(y)+I_2(y)I_3(y)}{4Z}
  \qquad\text{with}\quad y=\m^2\,,
\end{equation}
where
\begin{equation}
  Z=3I_0(y)^2-4I_1(y)^2+3I_2(y)^2\,.
\end{equation}
Analogously,
\begin{equation}
  \alangle-{\sum_n}'\frac{1}{(z_n^2+\m^2)^2}+4\Big({\sum_n}'\frac{1}{z_n^2+\m^2}\Big)^2\arangle
  =\frac{2I_0^2+I_1^2-2I_2^2+I_3^2-I_0I_2-2I_1I_3+I_2I_4}{8Z}\,.
\end{equation}
One can check that the massless sum rules (\ref{sum_1}) and
(\ref{sum_2}) are reproduced correctly in the chiral limit.  Many more
sum rules can be derived from the partition function for different
fermion masses, but we do not work them out explicitly. We only
mention that the sum rules corresponding to different masses satisfy
nontrivial consistency conditions as some of the masses are sent to
infinity, owing to decoupling \cite{Damgaard1998}.

\section{Conclusions}
\label{sec:conclusion}

In this paper we have constructed the low-energy effective Lagrangian
for two-color QCD with an even number of flavors at large quark
chemical potential $\mu$ based on the symmetry breaking pattern
induced by the formation of a diquark condensate.  For $N_f=2$, there
are two Nambu-Goldstone bosons $H$ and $\eta'$ associated with the
spontaneous breaking of the $\U(1)_B$ and $\U(1)_A$ symmetries,
respectively. The explicit breaking of $\U(1)_A$ vanishes at large
$\mu$ due to the screening of instantons.  For $N_f \geq 4$, there are
additional NG bosons (``pions'' $\pi_L$ and $\pi_R$) associated with
the chiral symmetry breaking pattern $\SU(N_f)_L \times \SU(N_f)_R
\rightarrow \Sp(N_f)_L \times \Sp(N_f)_R$.  We derived a mass formula
for the NG bosons in the case of degenerate quark masses.  
From the mixing of $\pi_L$ and $\pi_R$ by the mass term in the 
effective Lagrangian we found that half of them are exactly 
massless and that the other half are massive, with a mass
proportional to the quark mass as a consequence of the $\Z(2)_L \times
\Z(2)_R$ symmetry of the diquark pairing.

On the basis of the mass hierarchy near the Fermi surface, we have
identified a new finite-volume $\epsilon$-regime for the superfluid phase
at large $\mu$, where the mass scale of the non-NG modes is
characterized by the fermion gap $\Delta$.  In this regime, we can
exactly determine the quark mass dependence of the partition function
from the effective theory.  Matching this result against the two-color
QCD partition function, we have derived novel spectral sum rules for
inverse powers of the complex eigenvalues of the Dirac operator.  Our
sum rules explicitly show that the Dirac spectrum at large $\mu$ is
governed by the fermion gap $\Delta$, unlike the spectrum at low
$\mu$, which is governed by the chiral condensate as shown in
\cite{Smilga1995}.

A remarkable property of two-color QCD is that the
fermion sign problem is absent at nonzero $\mu$ 
(for even $N_f$ with pairwise degenerate quark masses)
because of the
pseudo-reality of $\SU(2)$.  Therefore, our sum rules can in principle
be checked in lattice QCD simulations.  This is in contrast to real
(three-color) QCD where the severe sign problem prevents us from
observing the presumed color superconductivity, although similar
spectral sum rules could be derived in the corresponding
$\epsilon$-regime \cite{Yamamoto2009}.  Testing our sum rules for two
colors on the lattice would enable us not only to verify the existence
of the BCS superfluid phase originating from diquark pairing, but also
to determine the value of the BCS gap $\Delta$ at large $\mu$
numerically.  This is important since previous studies of two-color
QCD at nonzero $\mu$ have only been able to determine the magnitude of
the diquark condensate, not the gap itself.

One should note that the diquark condensate
\emph{does not necessarily} imply a nonzero BCS gap $\Delta$.  At lower
$\mu$, just above $m_{\pi}/2$, the diquark condensate is a
strongly-coupled Bose-Einstein condensate (BEC) rather than the
weakly-coupled BCS pairing expected at large $\mu$ \cite{Hands2006},
although there is the possibility of a crossover between these two
states \cite{Splittorff2001} (see, however, \cite{Hands2006}).  
Also, the diquark condensate and the
BCS gap are intrinsically different in the context of the Dirac
spectrum in that the former (the latter) is governed by the radial
(the anisotropic two-dimensional) distribution of Dirac eigenvalues
\cite{Kanazawa2009}.  We are planning to embark on a lattice
simulation of two-color QCD at large $\mu$ to test the ideas proposed
in this paper and to make a first step towards providing a direct
signature of the BCS phase of two-color QCD (i.e., the gap around the
Fermi surface) at large $\mu$.

The BEC-BCS crossover, or the continuity between the
hadronic phase and the BCS superfluid phase \cite{Schafer:1998ef}, may
be closely related to the formal similarity of the partition functions
for degenerate quark masses at small $\mu$, see
\cite[(5.13)]{Smilga1995}, and at large $\mu$, see (\ref{eq:pf})
obtained in this paper.  In the case of real (three-color) QCD with
three flavors, it was recently suggested that the axial anomaly (or
the effect of instantons), which acts as an external field for the
chiral condensate, can lead to a crossover between the hadronic phase
and the color superconducting phase \cite{Hatsuda:2006ps,
  Yamamoto2007a, Yamamoto:2008zw} as an explicit realization of the
quark-hadron continuity in \cite{Schafer:1998ef}.  However, for QCD
with two colors and an even numbers of flavors, the axial anomaly
never acts as an external field for the chiral condensate, hence
another mechanism to account for the crossover phenomenon would be
necessary.

Finally, regarding the Dirac spectrum, it would be interesting to
obtain an analogue of the Banks-Casher relation at large $\mu$ and the
concrete form of the microscopic spectral density in the
$\epsilon$-regime we identified.  In particular, it is a challenging
problem to construct the corresponding random matrix model, which has
turned out to be very successful at $\mu=0$ and small $\mu$, to
reproduce our spectral sum rules and to elucidate universal properties
of the Dirac spectrum at large $\mu$ for both two and three colors.
(Earlier random matrix model approaches at large $\mu$ include
\cite{Vanderheyden2000,Vanderheyden2000a,Pepin2001,Vanderheyden2001,Klein2003,Klein2005}.)
The partial quenching technique (see, e.g.,
\cite{Damgaard1999,Akemann2004,Basile2007}) may well be relevant in
this regard.  A detailed analysis of these issues is deferred to future
work.

\acknowledgments

We thank Tetsuo Hatsuda, Shoichi Sasaki and Hiroshi Suzuki for helpful 
discussions and comments, and Thomas Sch\"afer and Jacobus Verbaarschot 
for clarifying communications. 
We are especially grateful to Tetsuo Hatsuda for 
a careful reading of the manuscript.
TK and NY are supported by the Japan Society for the 
Promotion of Science for Young Scientists. 
TW acknowledges support by JSPS and by the G-COE
program of the University of Tokyo and thanks the Theoretical Hadron 
Physics Group at the University of Tokyo for their hospitality while 
part of this work was being performed.

\appendix

\section{Microscopic derivation of mass terms in the effective
  theory}\label{app_A}
\label{sec:appendix}

In this appendix we give the detailed derivation of (\ref{c}) and
(\ref{c'}).  For this purpose, we calculate the shift of the vacuum
energy due to the quark mass from the microscopic theory (QCD), which
should be matched against the result obtained from the low-energy
effective Lagrangian (\ref{L}).  The starting point is the mass term
of the microscopic theory given in \cite{Schafer2002},
\begin{align}
  \mathcal{L}_\text{mass}= \frac{g^2}{8\mu^4}(\psi_{i,L}^{a
    \dag}C\psi_{j,L}^{b, \dag})(\psi_{k,R}^c C \psi_{l,R}^d)
  \left[(\tau^A)^{ac} (\tau^A)^{bd} M_{ik}M_{jl} \right] +
  (L\leftrightarrow R,\, M\leftrightarrow M^{\dag})\,,
  \label{eq:mass}
\end{align}
where $i,j,k,l$ ($a,b,c,d$) are flavor (color) indices and $\tau^{A}$
($A=1,\ldots,N_c^2-1$) are $\SU(N_c)$ color generators ($N_c=2$ here)
with normalization $\tr(\tau^A \tau^B)=\delta^{AB}/2$.  The
diquark condensates $d_L$ and $d_R$ that are antisymmetric in flavor
and color can be written as 
\begin{equation}
  \langle \psi_{i,L}^a C \psi_{j,L}^b \rangle = \epsilon^{ab} I_{ij} d_L\,,
  \qquad
  \langle \psi_{i,R}^a C \psi_{j,R}^b \rangle = \epsilon^{ab} I_{ij} d_R\,,
  \label{eq:d}
\end{equation}
where $I$ is defined in (\ref{I}). The expectation values of $d_L$ and
$d_R$ in the ground state can be evaluated in weak coupling.
Generalizing the result of \cite{Schafer2002} to arbitrary $N_c$, we
obtain
\begin{equation}
  |d_L|=|d_R|=\sqrt{\frac{6N_c}{N_c+1}}\frac{\mu^2 \Delta}{\pi g}\,.
  \label{eq:mag}
\end{equation}
Using (\ref{eq:d}), (\ref{eq:mag}), and the Fierz identity
\begin{equation}
  (\tau^A)^{ac}(\tau^A)^{bd}=\frac{1}{2} \delta^{ad}\delta^{bc}
  -\frac{1}{2N_c}\delta^{ac}\delta^{bd}\,,
\end{equation}
the shift of the vacuum energy due to the quark mass is given by
\begin{equation}
  \Delta \mathcal{E}= \frac{3\Delta^2}{4\pi^2} \tr(M I M^T I) +
  (M \leftrightarrow M^{\dag})\,.
  \label{eq:mass2}
\end{equation}
For $N_f=2$ we have $I_{ij}=\epsilon_{ij}$, and
(\ref{eq:mass2}) reduces to
\begin{equation}
  \Delta \mathcal{E}_{N_f=2}= -\frac{3\Delta^2}{2\pi^2} \det M +
  (M \leftrightarrow M^{\dag})\,.
  \label{eq:mass3}
\end{equation}
By comparing (\ref{eq:mass2}) or (\ref{eq:mass3}) with the vacuum
energy obtained from the low-energy effective Lagrangian (\ref{L})
(with $\Sigma_L=\Sigma_R=I$) or \eqref{c'}, the values of the
coefficients are found to be
\begin{equation}
  c=  \frac{3}{4 \pi^2}\,, \qquad c'= \frac{3}{2\pi^2}\,,
\end{equation}
respectively, where $c'$ corrects the value of $4/3\pi^2$
given in \cite{Schafer2003}.

\providecommand{\href}[2]{#2}\begingroup\raggedright\endgroup
\end{document}